\documentclass[prl,aps,twocolumn,superscriptaddress]{revtex4-1}
\usepackage{amsmath, amssymb, amsbsy}
\usepackage{color}
\usepackage{array}
\usepackage{bm, bbm}
\usepackage{graphicx}
\usepackage{diagbox, multirow}
\usepackage[percent]{overpic}
\usepackage{epsfig}
\usepackage{extarrows}
\usepackage{appendix}
\usepackage{txfonts}
\usepackage[normalem]{ulem}

\usepackage[colorlinks=true,linkcolor=blue,citecolor=blue,urlcolor=blue,bookmarks=false]{hyperref}

\begin{document}

\title{Unveiling the S=3/2 Kitaev Honeycomb Spin Liquids}
\date{\today}

\author{Hui-Ke Jin}
\affiliation {Department of Physics TQM, Technische Universit\"{a}t M\"{u}nchen,  James-Franck-Straße 1, D-85748 Garching, Germany}

\author{W. M. H. Natori}
\affiliation{Institute Laue-Langevin, BP 156, 41 Avenue des Martyrs, 38042 Grenoble Cedex 9, France}
\affiliation{Blackett Laboratory, Imperial College London, London SW7 2AZ, United Kingdom}

\author{F. Pollmann}
\affiliation {Department of Physics CMT, Technische Universit\"{a}t M\"{u}nchen,  James-Franck-Straße 1, D-85748 Garching, Germany}
\affiliation{Munich Center for Quantum Science and Technology (MCQST), 80799 Munich, Germany}

\author{J. Knolle}
\affiliation {Department of Physics TQM, Technische Universit\"{a}t M\"{u}nchen,  James-Franck-Straße 1, D-85748 Garching, Germany}
\affiliation{Munich Center for Quantum Science and Technology (MCQST), 80799 Munich, Germany}
\affiliation{Blackett Laboratory, Imperial College London, London SW7 2AZ, United Kingdom}

\begin{abstract}
The S=3/2 Kitaev honeycomb model (KHM) is a quantum spin liquid (QSL) state coupled to a static Z$_2$ gauge field. Employing an SO(6) Majorana representation of spin3/2’s, we find an exact representation of the conserved plaquette fluxes in terms of static $Z_2$ gauge fields akin to the S=1/2 KHM which enables us to treat the remaining interacting matter fermion sector in a parton mean-field theory.  We uncover a ground-state phase diagram consisting of gapped and gapless QSLs.  Our parton description is in quantitative agreement with numerical simulations, and is furthermore corroborated by the addition of a [001] single ion anisotropy (SIA) which continuously connects the gapless Dirac QSL of our model with that of the S=1/2 KHM. In the presence of a weak [111] SIA, we discuss an emergent chiral QSL within a perturbation theory.
\end{abstract}
\maketitle

{\bf\color{blue} Introduction.}
The search for quantum spin liquids (QSLs) has been at the forefront of condensed matter physics for many decades because they represent novel quantum phases of matter beyond the Landau paradigm of symmetry breaking --- instead they are characterized by fractionalized excitations and non-local quantum entanglement~\cite{Broholm2020,Knolle2019,Zhou2017,Savary2016}.
A paradigmatic example of a two-dimensional (2D) QSL is the seminal Kitaev honeycomb model (KHM)~\cite{Kitaev06}, which was initially derived to illustrate the basic ideas of topological quantum computation~\cite{NayakRMP}. Remarkably, the model has an exact solution which shows that its excitations are free Majorana fermions with a Dirac dispersion and gapped conserved plaquette fluxes which couple to the Majoranas via a \textit{static} Z$_2$ gauge field. In the context of frustrated magnetism research, the KHM provided a first rigorous example how a QSL with fractionalized excitations and emergent gauge fields can emerge in a concrete microscopic 2D spin model. 

In the last years, the KHM has transformed from a theoretical toy model to one of experimental relevance because a flurry of spin-orbit-coupled $4d$ and $5d$ transition metal compounds~\cite{Jackeli2009,Chaloupka2010,Kee2016,Trebst2022,Hermanns2018,Takagi2019} has been proposed as candidates for realizing its bond-anisotropic Ising interactions.
Remarkably, experiments have also observed signatures of the proximate Kitaev spin liquid (KSL) in several materials with effective spin 1/2 moments, such as $\alpha$-RuCl$_3$~\cite{Plumb2014,Sandilands2015,Banerjee2016,Do2017,Baek2017,Zheng2017} and (Na$_{1-x}$Li$_x$)$_2$IrO$_3$~\cite{Cao2013,Manni2014}, despite the residual zigzag ordered state which appears at low temperature ~\cite{Chaloupka2013,Yamaji2016} because of additional interactions, e.g. an off-diagonal symmetric $\Gamma$ exchange~\cite{Plumb2014}.
However, the exchange frustration of the KHM is not restricted to spin 1/2 and recently some promising realizations of higher-spin Kitaev materials have been proposed based upon $3d$ orbitals~\cite{XuNPJ2018,Xu2020,Stavropoulos2021}, in which the QSL-disrupting non-Kitaev exchanges might be reduced ~\cite{Liu2018,Sano2018,Liu2020}.
In particular, a microscopic derivation of the S=3/2 KHM model with an extra single ion anisotropy (SIA) has been established for the quasi 2D systems CrI$_3$ and CrGeTe$_3$~\cite{XuNPJ2018,Xu2020}.

After the original proposal of the S=1/2 KHM~\cite{Kitaev06}, much effort has been devoted to investigating the KHM models for S$>$1/2, which have not found an exact solution~\cite{Baskaran2007,Baskaran2008,Chandra2010,Oitmaa2018,Koga2018,Rousochatzakis2018,Dong2020,YBKim2020}.
Nevertheless, Baskaran {\em et al.}~\cite{Baskaran2008} showed early on that a generic spin-S KHM still has conserved Z$_2$ fluxes for each elementary hexagon and suggested via a semiclassical analysis that the ground state of KHMs for all values of S exhibits a homogeneous flux configuration in which all values of Z$_2$ fluxes are $+1$~\cite{Baskaran2008}. Subsequently, it has been proposed that the ground states of the S$>$3/2 KHMs are Z$_2$ QSLs described by an effective toric code on a honeycomb superlattice, but the employed semi-classical analysis breaks down precisely at S=3/2 ~\cite{Rousochatzakis2018}.
The S=1 KHM is amenable to numerical investigations and studies using exact diagonalization~\cite{Koga2018} and density matrix renormalization group (DMRG)~\cite{Dong2020} point to a gapless QSL ground state, whereas a tensor network approach proposes a gapped QSL for the isotropic model~\cite{YBKim2020}. Overall, the S=3/2 KHM seems to be the least understood of all Kitaev models --- the conserved plaquette fluxes alone do not help to gain an analytical understanding and a high density of low energy excitations lead to strong finite-size effects for numerical investigations.

Here, we report new exact properties of  the S=3/2 KHM and provide a systematic understanding of its ground-state phase diagram and excitations.
We introduce an SO(6) Majorana representation for the spin-3/2's which permits an exact mapping of the spin model to one of fermions coupled to a \textit{static} Z$_2$ gauge field. The latter determines the conserved plaquette flux just like in the original S=1/2 KHM. Within a given gauge field configuration, the Hamiltonian still contains quartic and even sextic fermion interaction terms but we construct a parton mean-field (MF) theory which turns out to be even quantitatively reliable. Our theory is furthermore corroborated by the addition of an extra [001] SIA to the S=3/2 KHM, which still preserves the conservation of the Z$_2$ fluxes and allows us to map out the phase diagram consisting of two gapless Dirac and two gapped QSLs. In the limit of large  [001] SIA, we find an exact solution of an effective S=1/2 KSL, which is continuously connected to one of the two gapless phases of the pure S=3/2 KHM. We also investigate the model using DMRG~\cite{White1992,White1993} and  find the numerical results to be in quantitative agreement with the predictions from our parton {MF theory}.  In the presence of a weak [111] SIA, we can derive an effective Hamiltonian within the zero-flux sector and argue that a chiral KSL is established.

\begin{figure}[t]
	\includegraphics[width=0.99\linewidth]{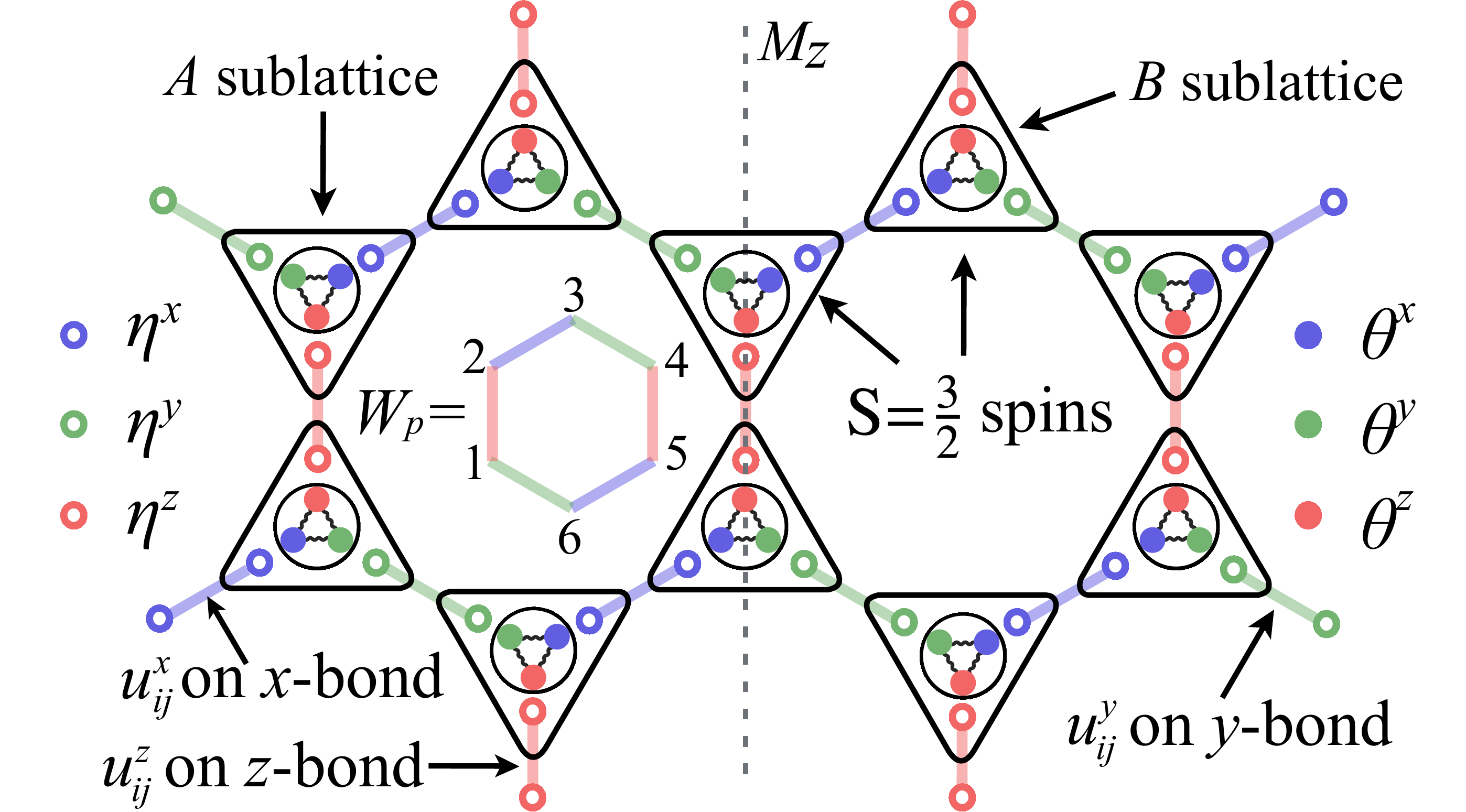}
	\caption{Graphic representation of the S=3/2 KHM~\eqref{eq:HMajorana} and the SO(6) Majorana representation.  Down (up) triangles stand for S=3/2 spins on $A$ ($B$) sublattice. Each spin is represented by SO(6) Majoranas, {\em e.g.}, three gauge Majoranas $\eta^{x,y,z}$ (circles) and three itinerant  Majoranas $\theta^{x,y,z}$ (dots). The blue, green, and red bold lines denote the \textit{static} Z$_2$ gauge fields $u^{x}$, $u^{y}$, and $u^{z}$ for $x$-, $y$-, and $z$-bond Ising interactions, respectively. The plaquette operator $W_p\equiv -e^{i\pi(S^x_1 + S^y_2 + S^z_3 + S^x_4 + S^y_5 + S^z_6)}$ can be expressed as the product of $u^a_{ij}$ around hexagon $p$. The gray dashed line stands for the mirror symmetry $M_z$ with $J_x=J_y$.}
	\label{fig:KitaevHoneycomb}
\end{figure}

{\bf\color{blue} Model Hamiltonian.}
The Hamiltonian of the S=3/2 KHM reads
\begin{align}
\mathcal{H}= \sum_{\langle ij\rangle_a} J_a S^a_i S^a_j, \label{eq:H1}
\end{align}
where $S_{i}^{a}$ $(a=x,y,z)$ are three components of an S=3/2 spin at site $i$ and $\langle{}ij\rangle_a$ denotes the nearest neighbor (NN) bonds of $a$-type S=3/2 Ising interactions (see Fig.~\ref{fig:KitaevHoneycomb}). 
There exist commuting plaquette operators $W_p$ for each hexagon $p$ (see Fig.~\ref{fig:KitaevHoneycomb}) as 
$W_p \equiv -e^{i\pi(S^x_1 + S^y_2 + S^z_3 + S^x_4 + S^y_5 + S^z_6)}$~\cite{Baskaran2008}.
By noticing that $[W_p, \mathcal{H}]=0$, the total Hilbert space of Hamiltonian \eqref{eq:H1} can be divided into orthogonal sectors characterized by flux configurations $\{w_p=\pm{}1\}$, where $w_p$ is the eigenvalue of $W_p$. 

{\bf\color{blue} SO(6) Majorana representation.} 
We introduce three gauge Majorana fermions $\eta^{a}_i$ and three itinerant Majorana fermions $\theta^a_i$ $(a=x,y,z)$, to obtain the SO(6) Majorana representation for spin-3/2's~\cite{Wang2009,Natori2016,Natori2017,Natori2018,Yao2009,Chua2011,Yao2011,Carvalho2018,Chulliparambil2020,Farias2020,Seifert2020,Natori2020,Ray2021,Chulliparambil2021}:
$S^a_i =  \frac{i}{4}\epsilon_{abc}\eta^b_i\eta^c_i - \frac{i}{2}\eta^a_i\tilde{\theta}^{a}_i,$
where $\tilde{\theta}^{x(y)}_i=\theta^z_i-(+)\sqrt{3}\theta^x_i$, $\tilde{\theta}^{z}_i=-2\theta^z_i$, and $\epsilon_{abc}$ is the Levi-Civita tensor (summation over repeated indices throughout).
This parton representation doubly enlarges the Hilbert space and the physical Hilbert space of spin-3/2's can be restored by imposing the local constraint $D_{i} = i\eta^x_i\eta^y_i\eta^z_i\theta^x_i\theta^y_i\theta^z_i=1$.
One can then obtain $i\eta^b_i\eta^c_i=\epsilon_{abc}\eta^a_i\theta^x_i\theta^y_i\theta^z_i$ and rewrite the spin operators as
\begin{align}
S^a_i = \frac{i}{2}\eta^a_i\left(\theta^{xyz}_i - \tilde{\theta}^{a}_i\right),~(a=x,y,z), \label{eq:Sa_M}
\end{align}
where $\theta^{xyz}_i=-i\theta^x_i\theta^y_i\theta^z_i$.

Eq.~\eqref{eq:H1} then becomes an effective Hamiltonian for Majoranas
\begin{align}
H=&-\frac{i}{4}\sum_{\langle{}ij\rangle_{a}}J_au^a_{ij}\left(\theta^{xyz}_i-\tilde{\theta}^{a}_{i}\right)	\left(\theta^{xyz}_j-\tilde{\theta}^{a}_{j}\right)\label{eq:HMajorana}, 
\end{align}
where $u^a_{ij}=i\eta^a_i\eta^a_j$. One can now verify that $[u^a_{ij},u^b_{kl}]=0$ for all different bonds and $[u^a_{ij}, H]=0$. Therefore, $u^a_{ij}$ with eigenvalues $\pm1$ is a {\em static} Z$_2$ gauge field!  Similar to the S=1/2 KHM, the plaquette operator $W_p$ is exactly mapped to a product of $u^a_{ij}$ around hexagon $p$, {\em e.g.}, $W_p=u^z_{12} u^x_{32}u^y_{34}u^z_{54} u^x_{56}u^y_{16}$  (see Fig.~\ref{fig:KitaevHoneycomb}).
In a fixed Z$_2$ gauge field configuration the Hamiltonian only depends on the itinerant Majoranas $\theta^{a}_i$. 

The microscopic derivation of the S=3/2 KHM introduced in Ref.~\cite{XuNPJ2018,Xu2020,Stavropoulos2021} suggests that it is usually accompanied by an extra [111] SIA term of the form $(S^c_i)^2$ with $S^c_i=\frac{1}{\sqrt{3}}\left(S^x_i+S^y_i+S^z_i\right)$.
In addition to the [111] SIA, we also consider a simplified [001] SIA term and focus on the Hamiltonian
\begin{align}
	\mathcal{H}_{\rm D}= \mathcal{H} + \sum_{i}\left[D_z(S^z_i)^2 + D_c\left(S^c_i\right)^2\right].
	\label{eq:HDz}
\end{align}
Note that $[W_p,\ \sum_{i} (S^z_i)^2]=0$ but the [111] SIA  breaks the conservation of fluxes, namely, $[W_p,\ \sum_{i} (S^c_i)^2]\neq{}0$. Therefore we always treat $D_c$ as a small perturbation to ensure that the system is  close to the Kitaev limit.

%\section{\em Parton construction.}
{\bf\color{blue} The [001] SIA limit.}
%{\bf Parton MF theory---}
First, we focus on the $D_c=0$ limit with conserved fluxes. Since the local S=3/2 states of $|S^z_i{\rm=}\pm\frac{3}{2}\rangle$ $\left(|S^z_i{\rm=}\pm\frac{1}{2}\rangle\right)$ will be energetically favored when $J_z$ ($D_z$) dominates, we expect that the competition of $J_z$ and $D_z$ leads to a rich phase diagram.  Moreover, below we show that for $D_z \to \infty$, we can recover an effective S=1/2 KSL.

\begin{figure}[t]
	\includegraphics[width=0.99\linewidth]{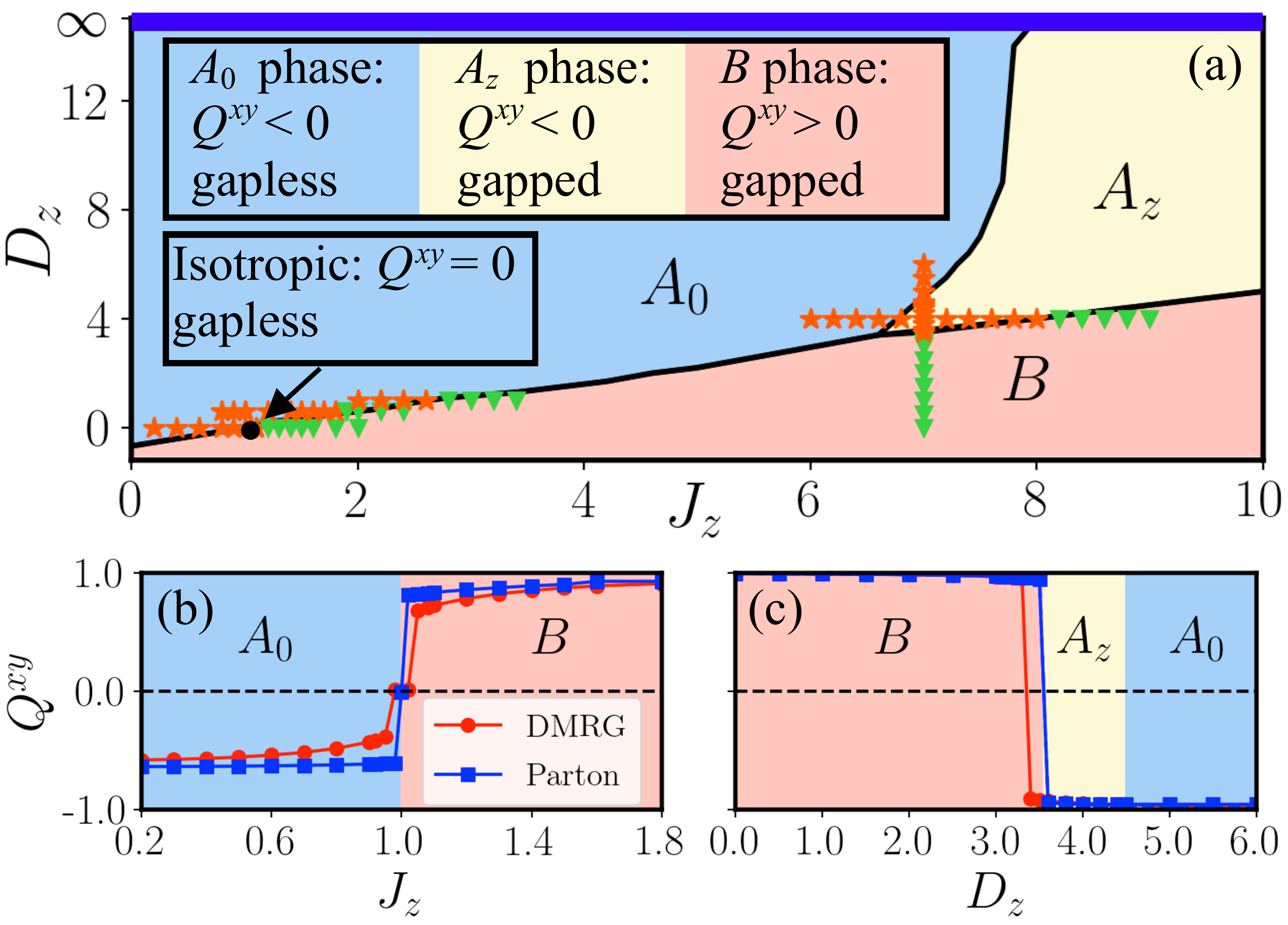}
	\caption{The ground-state phase diagram and the values of parameter $Q^{xy}$.
		(a) The MF phase diagram of Hamiltonian \eqref{eq:HMF} in zero-flux sector on a $36\times{}36$ torus.  The $A_0$ phase is a Dirac QSL with spin quadrupolar parameter $Q^{xy}<0$. In the $A_z$ ($B$) phase, the Majorana excitations are gapped with $Q^{xy}<0$ ($Q^{xy}>0$).  At the isotropic point, the ground state is a QSL with $Q^{xy}=0$.   The bold blue line at $D_z=\infty$ with $J_z<8$ ($J_z>8$) represents the effective gapless (gapped) S=1/2 KSL. The gapless phases in S=3/2 and S=1/2 KHMs can continuously connect to each other through the $A_0$ phase without energy gap opening. The orange stars (green triangles) represent the ground states obtained by DMRG on a $3\times{}4$ torus with zero-flux (disordered-flux) configurations with bond dimension $\chi=4000$.  (b, c) The values of $Q^{xy}$ as a function of $J_z$ with $D_z=0$  (b) and $D_z$ with $J_z=7$ (c). The computations of DMRG and parton MF theory are performed on a $3\times4$ torus.}\label{fig:phase_Qxy}
\end{figure}

By using $(S^z_i)^2 = -i\theta^x_i\theta^y_i$ (a constant of 5/4 has been omitted), the effective Majorana Hamiltonian can be divided into two parts. One is a quadratic Hamiltonian $H^{(2)}(\{u\})\equiv\frac{-i}{4}\sum_{\langle{}ij\rangle_{a}}J_au^a_{ij}\tilde{\theta}^a_i\tilde{\theta}^a_j-iD_z\sum_i\theta^x_i\theta^y_i$, and the other is an interacting Hamiltonian consisting of quartic and sextic terms, which we treat within a MF analysis. 
In our decoupling scheme, the MF Hamiltonian reads %can be written in a compact form as 
\begin{align}
H_{\rm MF}(\{u\})&=H^{(2)}(\{u\})+\sum_{\langle{}ij\rangle_{a}}\frac{iJ_au^a_{ij}}{4}\Bigg\{\frac{\epsilon_{opq}\epsilon_{rst}}{4}\langle\theta^{o}_i\theta^{p}_i\theta^{r}_j\theta^{s}_j\rangle{}\theta^{q}_i\theta^{t}_j \notag\\
+&\frac{\epsilon_{lmn}}{2}\left(\theta^{m}_i\theta^{n}_i\langle\theta^{l}_i\theta^x_j\theta^y_j\theta^z_j\rangle-\theta^{m}_j\theta^{n}_j\langle\theta^{l}_j\theta^x_i\theta^y_i\theta^z_i\rangle\right)\label{eq:HMF}\\
+&\frac{\epsilon_{uvw}}{2}\left[\left(Q^{uv}_{i}\theta^{w}_i\tilde{\theta}^{a}_j-\Delta^{w\tilde{a}}_{ij}\theta^{u}_i\theta^{v}_i+iQ^{uv}_{i}\Delta^{w\tilde{a}}_{ij}\right)-(i\leftrightarrow j)\right] \Bigg\} \notag
\end{align}
with the on-site parameters $Q^{ab}_{i}\equiv{}-\langle{}i\theta^{a}_i\theta^b_i\rangle$ ($a\neq{}b$) and bond parameters $\Delta^{ab}_{ij}\equiv{}\langle{}i\theta^{a}_i\theta^b_j\rangle$ $\left(\Delta^{a\tilde{b}}_{ij}\equiv{}\langle{}i\theta^{a}_i\tilde{\theta}^b_j\rangle\right)$. 
In accordance with Wick's theorem, the average of quartic terms can be decoupled as $\langle\theta^{o}_i\theta^{p}_i\theta^{r}_j\theta^{s}_j\rangle=-Q^{op}_{i}Q^{rs}_{j}+\Delta^{or}_{ij}\Delta^{ps}_{ij}-\Delta^{os}_{ij}\Delta^{pr}_{ij}$ and $\langle\theta^{l}_i\theta^{x}_j\theta^{y}_j\theta^{z}_j\rangle=\Delta^{lx}_{ij}Q^{yz}_{j}+\Delta^{ly}_{ij}Q^{zx}_{j}+\Delta^{lz}_{ij}Q^{xy}_{j}$. 
Eventually, $H_{\rm{MF}}$ is parameterized by MF parameters $Q$ and $\Delta$, which we determine self-consistently (see Supplementary Note 1). In the presence of Z$_2$ flux conservation, we can restrict to antiferromagnetic couplings of $J_a>0$ since in the parton level a sign change of $J_a\rightarrow{}-J_a$ can be resolved by $u^a_{ij}\rightarrow{}-u^a_{ij}$. 
For simplicity, we focus on the case of $J_x=J_y=1$ with a mirror symmetry $M_z$ across the $z$-bonds (see Fig.~\ref{fig:KitaevHoneycomb}) which leads to a vanishing of $Q^{yz,zx}=0$. The only nonzero on-site parameter is $Q^{xy}\equiv-\langle{}i\theta^x\theta^y\rangle=\langle(S^z)^2\rangle$ which is a time-reversal-invariant spin quadrupolar component characterizing the MF ground states of $H_{\rm MF}(\{u\})$ (see Supplementary Note 2).

Following the proposal of Ref.~\cite{Baskaran2008} that for generic spin-$S$ KHMs the ground-state always exhibits a zero-flux configuration with $\{w_p\}=1$, we will mainly focus on the zero-flux sector and fix $\{u\}=1$ for studying the  Majorana excitations of $H_{\rm MF}$.  Note that the quadratic Hamiltonian $H^{(2)}(\{u\})$ alone always energetically favors a zero-flux state.

\begin{figure}[t]
	\includegraphics[width=0.99\linewidth]{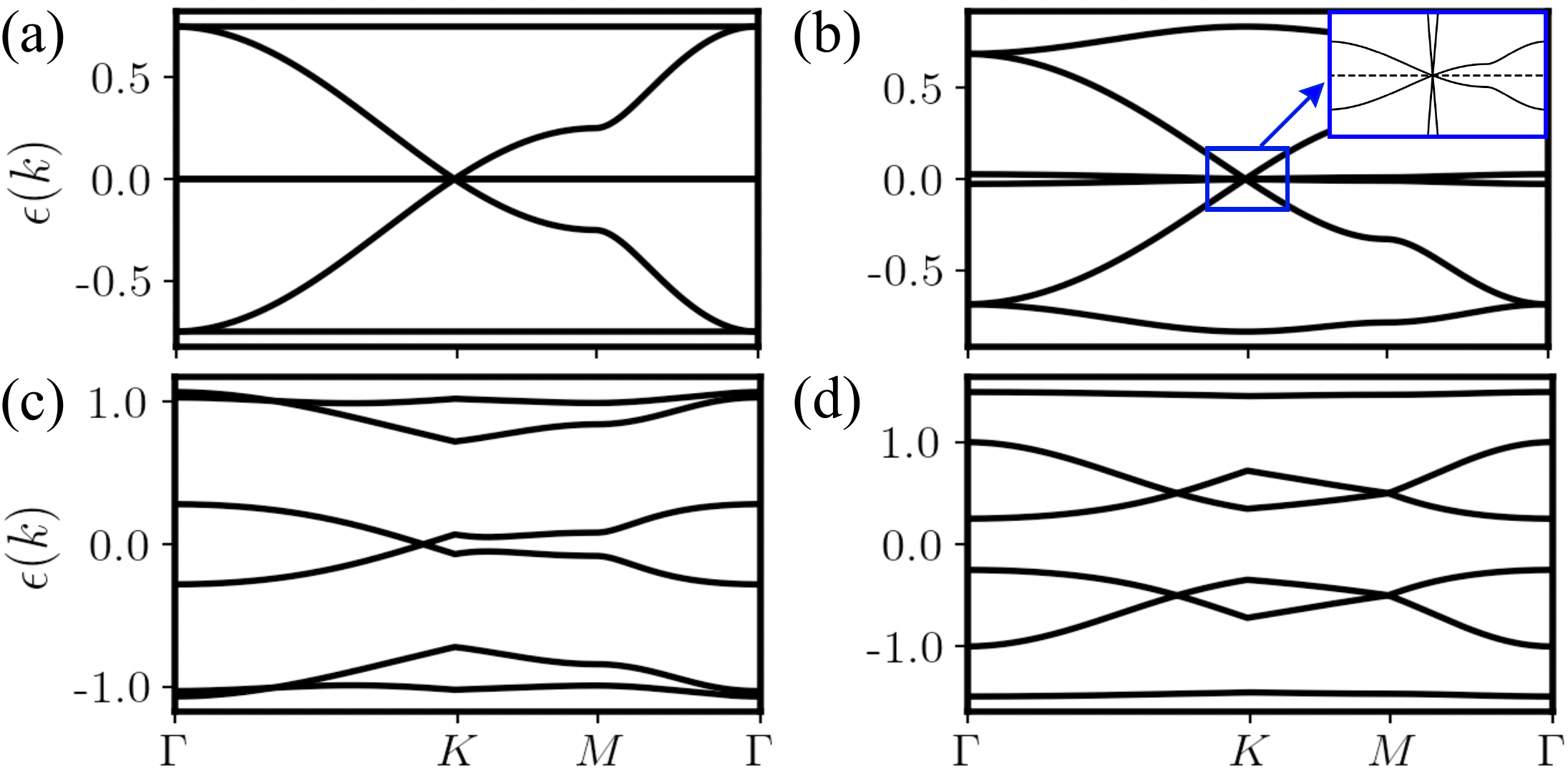}
	\caption{ Band structure of itinerant Majorana fermions. (a) The Majorana band structure of the quadratic Hamiltonian $H^{(2)}$ at the isotropic point $J_z=1$ and $D_z=0$. (b-d) The Majorana band structures of the MF Hamiltonian $H_{\rm MF}$ with (b) $J_z=1$ and $D_z=0$, (c) $J_z=1.2$ and $D_z=0.4$, and (d) $J_z=1.4$ and $D_z=0$. Here $J_x=J_y=1$,  and a zero-flux configuration of $\{u\}=1$ is used. The inset in (b) shows the zoom-in around the $K$ point.} \label{fig:bands}
\end{figure}

The Hamiltonian $H_{\rm MF}(\{u\}=1)$ displays four phases which are characterized by their distinct  Majorana excitations and the value of the quadrupolar parameter $Q^{xy}$, as shown in Fig.~\ref{fig:phase_Qxy}(a).
(i) At the isotropic point of $J_z=1$ and $D_z=0$, the ground state is a Dirac QSL with $Q^{xy}=0$.
The  Majorana band structure of $H_{\rm MF}(\{u\}=1)$ at the isotropic point is almost the same as that of the quadratic Hamiltonian $H^{(2)}(\{u\}=1)$, except that the exact flat bands populated by $\theta^y$ in Fig.~\ref{fig:bands}(a) acquire a very narrow dispersion as shown in Fig.~\ref{fig:bands}(b). 
Consequently, in this phase $H_{\rm MF}(\{u\}=1)$ has a Dirac point at the $K$-point, around which there exist two gapless  Majorana bands crossing linearly at zero energy but one of whose velocity is close to zero. 
(ii) In the $A_0$ phase, the ground state is a Dirac QSL coexisting with a spin quadrupolar parameter $Q^{xy}<0$.  Because of a nonzero $Q^{xy}<0$, only one branch of gapless spin excitations remains around the Dirac point [see Fig.~\ref{fig:bands}(c)]. 
(iii) In the $A_z$ phase, the effect of $D_z$ remains, but a relatively large anisotropy of $J_z$ gaps out all  Majorana excitations and the ground state is a gapped QSL coexisting with a spin quadrupolar parameter $Q^{xy}<0$.
(iv) In the $B$ phase, the dominant  $J_z$ leads to a positive $Q^{xy}>0$ and all Majorana excitations are gapped. Fig.~\ref{fig:bands}(d) shows a typical Majorana band structure for the $B$ phase.
We conclude that the transition between the $B$ phase and the $A_0$ ($A_z$) phase is of first-order since $Q^{xy}$ shows a discontinuous bump at the phase boundary in Fig.~\ref{fig:phase_Qxy}(b) and (c).

{\bf\color{blue}  Effective S=1/2 KSL.} A key advantage of using the [001] SIA is that the Hamiltonian can be solved exactly in the limit of $D_z\rightarrow\infty$, because the high energy local states of $|S^z_i=\pm{}\frac{3}{2}\rangle$ are removed and the ground-state subspace of Hamiltonian \eqref{eq:HDz} is spanned by the local states of $|S^z_i=\pm1/2\rangle$ only.
Within the framework of the SO(6) Majorana representation, the itinerant Majoranas $\theta^x_i$ and $\theta^y_i$ are paired up subjected to the constraint $i\theta^x_i\theta^y_i=1$ ($Q^{xy}=-1$). Then the three S=3/2 operators in Eq.~\eqref{eq:Sa_M} projected onto the subspace of $i\theta^x_i\theta^y_i=1$ become
\begin{align}
S^z_i\simeq \frac{i}{2}\eta^z_i\theta^z_i,\qquad{}S^{x(y)}_i\simeq{}i\eta^{x(y)}_i\theta^z_i\label{eq:4Majorana}.
\end{align}
Obviously, Eq.~\eqref{eq:4Majorana} is equivalent to Kitaev's original four-Majorana representation~\cite{Kitaev06}. 
It is remarkable that an effective S=1/2 KHM with a renormalized coupling constant $J_z\rightarrow{}J_z/4$ emerges in our S=3/2 Hamiltonian~\eqref{eq:HMajorana} for $D_z\rightarrow{}\infty$ with an effective gapless (gapped) S=1/2 KSL for $J_z<(>)~8$. This connection gives additional support to the assumption of a zero-flux ground state and our MF treatment, which is known to exactly capture the phase diagram and nature of excitations of the S=1/2 KHM~\cite{Burnell2011,Knolle2018A}. 
Indeed, we find that the gapless KSL of the emergent S=1/2 KHM is continuously connected to the gapless $A_0$ phase of the pure S=3/2 KHM in Fig.~\ref{fig:phase_Qxy}(a). 

{\bf\color{blue} Numerical results.} Since Lieb's theorem~\cite{Lieb1994} may not be applicable for the interacting Hamiltonian~\eqref{eq:HMajorana}, we examine the zero-flux ground-state configuration, which is a pivotal assumption in our parton theory.
In order to examine the reliability of our parton MF theory in the case of a [001] SIA,  we employ state-of-the-art DMRG method~\cite{White1992,White1993} to study the ground state of Hamiltonian~\eqref{eq:HDz}.
%DMRG is a very powerful numerical approach for studying 1D strongly-correlated systems.

To perform DMRG calculations on a 2D honeycomb lattice of $L_1\times{}L_2$ unit cells, we consider a cylindrical geometry for which the periodic boundary condition (PBC) is imposed along the shorter direction ({\em e.g.}, the circumference $L_1$), while the longer ({\em e.g.}, the length $L_2$) is left open.
Moreover,  we also adopt small tori with PBCs along both directions to strictly preserve the $A$/$B$ sublattice and translational symmetries. 
%If $\mathcal{H}_{\rm Dz}$ is placed on cylinders, the unpaired gauge Majoranas $\eta^{a}$ at each boundaries do not enter into $H_{\rm Dz}$ and thus lead to extra ground-state degeneracies.
The DMRG simulations are performed on a $3\times{}4$ torus as well as a $4\times{}8$ cylinder ($L_1=4$). %until the measured quantities are faithfully converged.
The bond dimension of DMRG is kept as large as $\chi=4000$, resulting in a typical truncation error of $\epsilon\simeq{}10^{-6}$ ($\epsilon\simeq{}10^{-4}$ close to the isotropic point $J_z=1$ and $D_z=D_c=0$). In general, we encounter significant finite-size effects in our numerical studies in contrast to checks on the S=1/2 and S=1 KHMs.

We find that the ground states in our DMRG  simulations always exhibit a zero-flux configuration in the $A_0$ and $A_z$ phases. %for both torus and cylindrical geometries.
In contrast, for the $B$ phase, DMRG does not converge to a unique ground-state flux configuration but instead leads to a disordered-flux ground state in which the measured flux for each plaquette is neither $1$ nor $-1$.  
The data points for different DMRG-obtained ground-state flux configurations are shown in Fig.~\ref{fig:phase_Qxy}(a).

The disordered-flux ground states obtained by DMRG indicate an extremely small flux gap above the zero-flux state in the $B$ phase. 
This flux gap can also be evaluated within our {MF theory} in the $B$ phase. To connect to our DMRG simulations, we have performed the MF calculations on the same $3\times{}4$ torus.
We find that the energy of a pair of neighboring fluxes is  $E_{\rm 2fluxes}\simeq{}10^{-6}$ for $J_z=1.2$ and $D_z=0$, which is as small as the corresponding DMRG truncation error $\epsilon\simeq{}10^{-6}$.
The flux gap $E_{\rm 2fluxes}$ rapidly decreases as $J_z$ increases, which explains why DMRG fails to capture the conserved Z$_2$ flux in the $B$ phase.

Next, we investigate the local spin quadrupolar parameter $Q^{xy}_{i}$ using DMRG and find that the $Q^{xy}_{i}$ are spatially uniform (on the torus and in the bulk of cylinders) as assumed in our parton description.
A remarkable observation is that, if we ignore small discrepancies near the phase boundaries, the values of the DMRG-obtained $Q^{xy}$ are even in quantitative agreement with those given by the parton {MF theory} within the zero-flux sector, as shown in Fig.~\ref{fig:phase_Qxy}(b) and (c).
%Notice that the DMRG-obtained values of $Q^{xy}$ are spatially uniform, which is also consistent with our {MFT}. %In Fig.~\ref{fig:phase_Qxy} we show the average values of $Q^{xy}$. 
For the purpose of comparison, we have chosen the same lattice geometry, {\em e.g.}, a $3\times{}4$ torus, to calculate the MF and DMRG values of $Q^{xy}$.
The difference between the MF $Q^{xy}$ for a $3\times{}4$ torus and its extrapolation to an infinite lattice is of the magnitude of $10^{-2}$ (except near the phase boundaries), thus finite-size effects for $Q^{xy}$ are small.
This remarkable quantitative agreement demonstrates that our SO(6) Majorana {MF theory} provides a compelling scenario for describing the S=3/2 KHM. For example, we can now understand why the S=3/2 KHM at the isotropic point is so challenging for numerical methods, {\em e.g.} the extreme system-size dependence encountered in our DMRG simulations, because the almost flat Majorana bands, see Fig.~\ref{fig:bands}(b), lead to a large pile up of close to zero energy states.

{\bf\color{blue} The [111] SIA limit.}
Next, we study the experimentally more relevant case, i.e., $J_x=J_y=J_z=J$, $D_z=0$, and $D_c\ll{}|J|$. A finite [111] SIA breaks the flux conservation, leading to a dynamical gauge field. In analogy to the procedure of treating a magnetic field in the S=1/2 KHM~\cite{Kitaev06}, we circumvent this problem by deriving an effective three-body quadrupolar interaction within the zero-flux sector. This can be further motivated by noticing that 
$(S^c_i)^2=-\frac{i}{3}\left(\eta_{i}^{x}+\eta_{i}^{y}+\eta_{i}^{z}\right)\theta_{i}^{y}$ plays a similar role as the [111] magnetic field in the $S=1/2$ KHM~\cite{Kitaev06}. %However, since the itinerant Majoranas $\theta^{y}$ are mainly constrained to the flat bands, we expect that the SIA effect will be much more drastic than that of the [111] magnetic field in S=1/2 KHM.
\begin{figure}[t]
	\includegraphics[width=0.99\columnwidth]{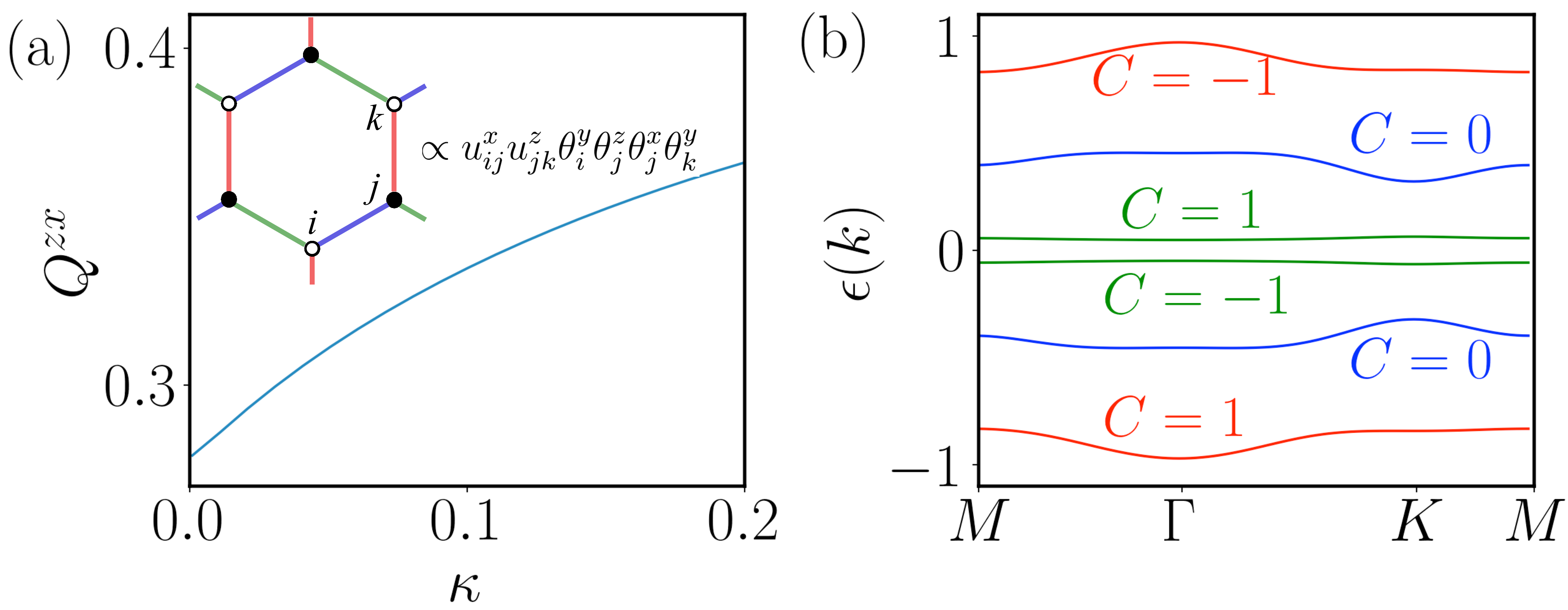}
	\caption{\label{fig:Qy_against_kappa} {Chiral QSL induced by [111] SIA.} (a) $Q^{zx}$ as a function of $\kappa$. Inset: A sketch for the three-body quadrupolar interactions in Eq.~\eqref{eq:H3_Majorana}. The blue, green, and red bonds stand for the $x$-, $y$-, and $z$-type S=3/2 Ising interactions, respectively. The summation in Eq.~\eqref{eq:H3_Majorana} takes place over the triangle with vertexes $i$, $j$, and $k$ (and symmetry-equivalent ones).   (b) The band structure and corresponding Chern numbers for the parton MF theory in Eq.~\eqref{eq:H3_MFT} with $\kappa=0.01$.}
\end{figure}

The effective term $H^{(3)}$ described by three-body
quadrupolar interactions is represented by 
\begin{align}
	H^{(3)} & =\kappa\sum_{\left\langle ij\right\rangle _{a}\left\langle jk\right\rangle _{b}}u^a_{ij}u^b_{jk}\left(i\theta_{i}^{y}\theta_{j}^{z}\right)\left(i\theta_{j}^{x}\theta_{k}^{y}\right),\label{eq:H3_Majorana}
\end{align}
where $\kappa\sim D_{c}^{3}/\Delta_{\text{flux}}^{2}$ ($\Delta_{\text{flux}}\approx{}0.093J$) and $\langle{}ij\rangle_{\alpha}$ and $\langle{}ij\rangle_{\beta}$ are two NN bonds connected by site $j$ [see Fig.~\ref{fig:Qy_against_kappa}(a)]. Eq.~\eqref{eq:H3_Majorana} clearly commutes with the Z$_2$ gauge fields but is quartic in the itinerant Majoranas. Its most general decoupling reads
\begin{align}
	H_{\text{MF}}^{(3)}(\{u\}=1)&
	= \kappa\sum_{\left\langle ij\right\rangle _{a}\left\langle jk\right\rangle_{b}}
	i\left(\Delta^{yz}_{ij}\theta^x_j\theta^y_k+\Delta^{yx}_{kj}\theta^z_j\theta^y_i\right)\nonumber \\
	&+\kappa\sum_{\left\langle ij\right\rangle _{a}\left\langle jk\right\rangle_{b}} i \left(-Q^{zx}_j \theta^y_i\theta^y_k + \xi^{yy}_{ik}\theta^x_j\theta^y_j\right), \label{eq:H3_MFT}
\end{align}
where $\xi^{yy}_{ik}=-\langle{}i\theta^y_i\theta^y_k\rangle$ is the hopping parameter on the 2$^{\text{nd}}$ NN bond. A nonzero $Q^{zx}$ breaks time-reversal symmetry (TRS) so that $H_{\text{MF}}^{(3)}$ naturally describes a chiral KSL analogous to the $S=1/2$ case~\cite{Kitaev06}.
A key difference is that here TRS is \emph{spontaneously}
broken since Eqs.~\eqref{eq:HDz} and \eqref{eq:H3_Majorana} are even under time-reversal.

$H_{\text{MF}}^{(3)}$ favors the $S=3/2$ chiral KSL, which undergoes a first-order phase transition on the parton MF level. For small $\kappa=0.01$, the self-consistent solution converges to parameters $Q^{zx}\approx0.28$ and $|\xi^{yy}|\approx0.115$. Fig.~\ref{fig:Qy_against_kappa}(a) presents the evolution of $Q^{zx}$ as a function of positive $\kappa$ --- solutions for negative $\kappa$ are obtained by changing the signs of $Q^{zx}$, $\xi^{yy}$, and $\Delta^{yx(z)}$. 

Next, we study properties of this chiral KSL. The Majorana hybridization induced by a nonzero $Q^{zx}$ narrows all dispersions and separates the six bands from each other, see Fig. \ref{fig:Qy_against_kappa}(b).
We evaluate the topological characteristics of these bands in terms of the Chern number, which is 
$$C\text{'s}=\left(1,0,-1,1,0,-1\right)$$
for the lowest to the highest band. Notice that in contrast to the $S=1/2$ chiral KSL, the sum of Chern numbers over the negative energy bands is zero. Therefore, no quantized  thermal Hall conductivity is expected in the low temperature limit, but an activated signal emerges for increasing temperatures.

{\bf\color{blue} Discussion.}
In summary, we have studied the ground-state phase diagram and excitations of the S=3/2 KHM with additional SIA terms. We employed a parton theory based on the SO(6) Majorana representation of spin=3/2's, which is supported by DMRG simulations. 
We have shown that the conserved flux for each honeycomb plaquette can be represented exactly via a static Z$_2$ gauge field similar to the well-known S=1/2 KHM, which is key for identifying the correct MF decoupling of the parton description. For a [001] SIA, DMRG calculations are shown to agree with our self-consistent {MF theory}  qualitatively and even quantitatively. We uncover a rich phase diagram characterized by distinct Majorana excitations and different phases with spin quadrupolar parameter $Q^{xy}=\langle(S^z)^2\rangle$: (i) a gapless Dirac QSL with $Q^{xy}=0$ and an additional almost flat Majorana band close to zero energy at the isotropic point ($J_z=1, D_z=0$); (ii) a gapless Dirac QSL with $Q^{xy}<0$ in the $A_0$ phase; (iii)  a gapped QSL with $Q^{xy}<0$ in the $A_z$ phase; and (iv) a gapped QSL with $Q^{xy}>0$ in the $B$ phase. For a dominating [001] SIA, the low energy sector of the S=3/2s reduces to effective S=1/2s which allows us to continuously connect the gapless $A_0$ phase of the pure S=3/2 KHM to that of the well-known Dirac QSL of the S=1/2 KHM.
In the $B$ phase, we found that DMRG fails to capture the conservation of Z$_2$ fluxes because of an extremely small flux gap above the zero-flux ground state, which is again accounted for in our parton MF theory.
In the presence of a small [111] SIA, we establish an effective model in the zero-flux sector with three-body quadrupolar interactions. Our parton MF study indicates an emergent chiral KSL spontaneously breaking TRS.

We argue that our SO(6) Majorana parton theory efficiently describes the different QSLs of the S=3/2 KHM, which also provides a compelling scenario for explaining the difficulties encountered in the numerical studies.
Hence, it will provide a good starting point for studying the robustness of the QSL regimes with respect to additional terms in the Hamiltonian, for example different exchange interactions and the SIA relevant for microscopic realizations of the S=3/2 KHM~\cite{Xu2020}. The connection to the S=1/2 KHM indicates that in particular the ferromagnetic QSLs will be very fragile and, in general, the formation of conventional magnetic order will be further facilitated because of flux-fermion bound state formation involving the almost flat Majorana bands. In that context,  large-scale numerical studies for S=3/2 KHM with non-Kitaev interactions like [111] SIA and Heisenberg terms are still highly demanded, and the quality of our parton MF states can be further improved by efficient tensor network representation with Gutzwiller projection~\cite{Jin2020,Jin2020_2,Tu2020} or by including different flux sectors in the variational ansatz~\cite{zhang2021variational}.   

In the future, it will be worthwhile to study the effect of applying a magnetic field and the ensuing QSL phases of the S=3/2 KHM. 
Similarly, it would be desirable to generalize our Majorana parton construction to higher-spin systems whose dimension of local Hilbert space is $2^n$ (with $n$ integer), {\em i.e.} the S=7/2 KHM could possibly have a similar exact static Z$_2$ gauge field permitting an efficient description via an eight Majorana representation for spin-7/2's.

{\bf\color{blue} Acknowledgement.}
F. P. acknowledges the support of the Deutsche Forschungsgemeinschaft (DFG, German Research Foundation) under Germany’s Excellence Strategy EXC-2111-390814868 and TRR 80. H.-K. J. is funded by the European Research Council (ERC) under the European Unions Horizon 2020 research and innovation program (grant agreement No. 771537). W. M. H. N. and J. K.  acknowledge the support from the Royal Society via a Newton International Fellowship through project NIF\textbackslash R1\textbackslash 181696.

\bibliography{S3_2KitaevSO6MF.bib}

\clearpage
\begin{widetext}
	\begin{Large}
\centering{Supplementary Information for ``Unveiling the S=3/2 Kitaev Honeycomb Spin Liquids''}
	\end{Large}
\renewcommand{\theequation}{\arabic{equation}}
\renewcommand{\thefigure}{\arabic{figure}}
\renewcommand{\figurename}{Supplementary Figure}
\renewcommand\refname{Supplementary Reference}
\renewcommand{\tablename}{Supplementary Table}
\renewcommand{\thetable}{\arabic{table}}

In this Supplementary Information, we show more details about (i) the self-consistent mean-field theory for S=3/2 Kitaev honeycomb model and (ii) the spin quadrupolar parameter.

\subsection*{Supplementary Note 1: Self-consistent mean-field theory}

In the main text, we have introduced a mean-field Hamiltonian $H_{\rm }(\{u\}=1)$ with the following mean-field  order parameters:
\begin{align}
	Q_{i}^{ab} & =-\left\langle i\theta_{i}^{a}\theta_{i}^{b}\right\rangle\ \  (a\neq{}b) ,\nonumber \\
	\Delta_{ij}^{ab} & =\left\langle i\theta_{i}^{a}\theta_{j}^{b}\right\rangle.
\end{align}
%in which we will be using the convention $\epsilon_{xyz}=1$ and $i(j)\in{}A(B)$ sublattice for nearest neighbor bond $\langle{}ij\rangle$.
%Notice that there is no order parameter involving $\eta$ fermions, since they serve as the static $Z_2$ gauge fields $u_{ij}^a=i\eta^{a}_i\eta^a_j=\pm{}1$ on the $a-$type nearest neighbor bonds $\langle{}ij\rangle{}_a$. 
For concreteness, we investigated the mean-field theory of the model with fixed exchange parameters $J_{x}=J_{y}=1$ and varying
$J_{z}$ and $D_z$. In this parameter regime, the model preserves mirror
symmetry $M_z$, inversion symmetry $I$, and time-reversal symmetry $\mathcal{T}$ (see Supplementary Fig.~\ref{fig:app_lattice}), which will impose constraints to the order parameters and allow us to provide a succinct form for the mean-field Hamiltonian $H_{\rm }(\{u\}=1)$. 
In accordance with a full symmetry analysis~\cite{Willian2021}, we find that for a translational invariant solution which preserves $M_z$, $I$, and $\mathcal{T}$ symmetries there exist only eight non-zero and independent mean-field order parameters
\begin{equation*}
	Q^{xy}, \Delta^{zx}_x, {\rm \ and\ } \Delta^{aa}_{z(x)},~ (a=x,y,z), %\Delta^{xy}_z, \Delta^{xy}_x, \Delta^{yz}_x, 
\end{equation*}
where the bond parameters $\mathrm{\Delta}_{x(z)}^{ab}$ are defined on the $x$-type ($z$-type) bonds.

\begin{figure}[th]
	\centering
	\includegraphics[scale=0.2]{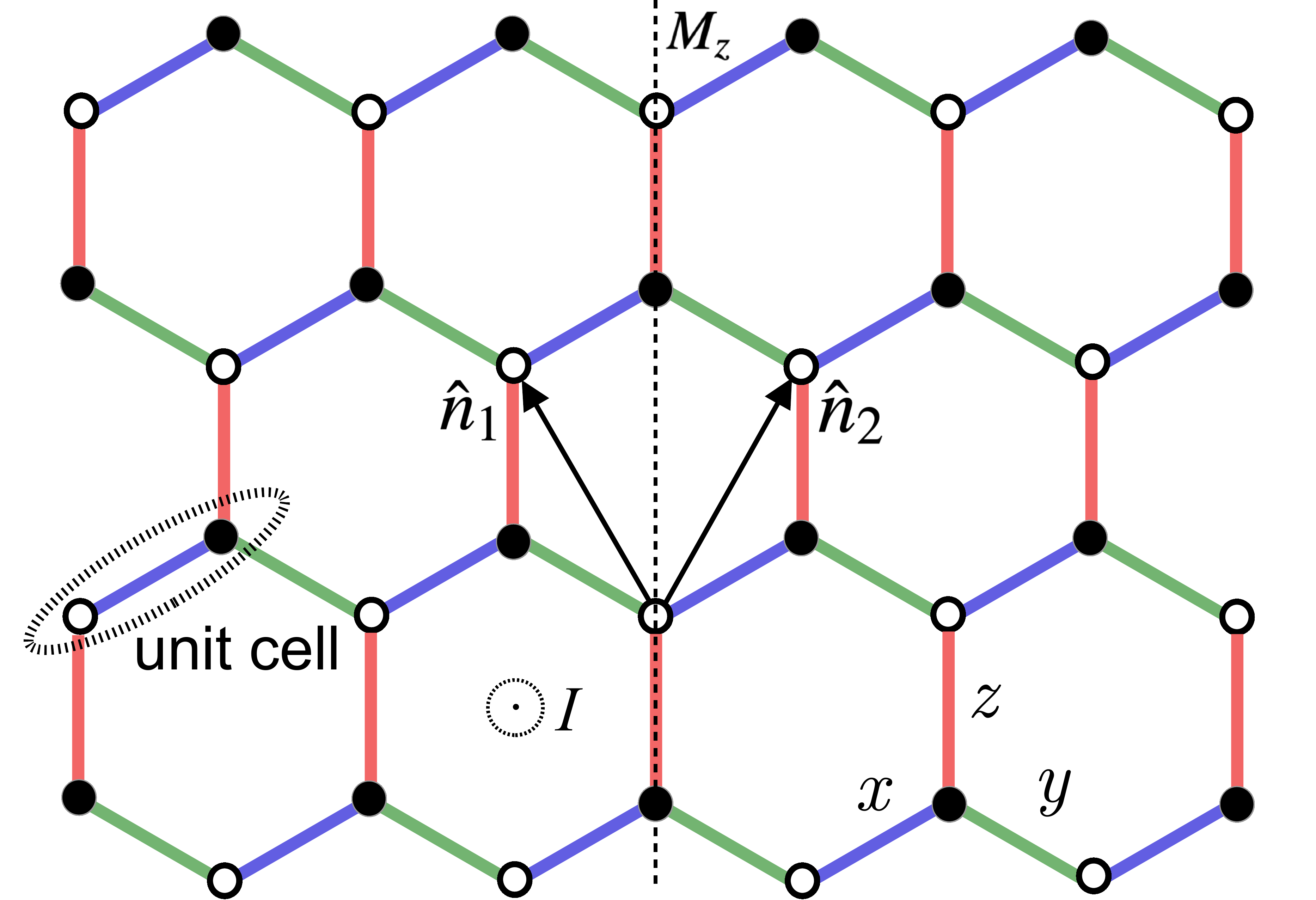}
	\caption{The mirror symmetry $M_z$ and inversion symmetries $I$ for the Kitaev model on a honeycomb lattice. $\hat{n}_1$ and $\hat{n}_2$ are two Bravais lattice vectors. The dots and circles denote the $A$ and $B$ sublattices, respectively.\label{fig:app_lattice}}
\end{figure}

We can write down the mean-field Hamiltonian $H_{\rm MF}(\{u\}=1)$ in the reciprocal space as
\begin{equation}
	H_{\rm MF}(\{u\}=1) = \sum_{\bf{k}}\psi_{\bf{k}}^\dagger{}H_{\bf{k}}\psi_{\bf{k}},\quad \psi^\dagger_{\bf{k}}=\left(\theta^z_{A,\bf{k}},\theta^x_{A,\bf{k}},\theta^y_{A,\bf{k}},\theta^z_{B,\bf{k}},\theta^x_{B,\bf{k}},\theta^y_{B,\bf{k}}\right)
\end{equation}
where the Fourier transformed fermions $$\theta^{a}_{A(B),\bf{k}}=\frac{1}{\sqrt{N}}\sum_{\bf{r}}e^{-i{\bf{}k}\cdot\bm{r}}\theta^a_{A(B),\bf{r}},~~\left(\theta^{a}_{A(B),\bf{k}}\right)^\dagger=\theta^{a}_{A(B),-\bf{k}}$$ on the $A~(B)$ sublattice are complex fermions rather than Majorana fermions, and $\bf{r}$ denotes the unit cell coordinates.
Then, the self-consistent equations can be written as 
\begin{subequations}
	\begin{align}
		& Q^{ab} = \frac{1}{N}\sum_{\bf{k}}-\left\langle{}i\theta^a_{A(B),\bf{k}}\theta^b_{A(B),-\bf{k}}\right\rangle,\\
		& \Delta^{ab}_{c}=\frac{1}{N}\sum_{\bf{k}}e^{-i{\bf{}k}\cdot\hat{n}_{c}}\left\langle{}i\theta^a_{A,\bf{k}}\theta^b_{B,-\bf{k}}\right\rangle{}, (c=x,y,z),
	\end{align}
\end{subequations}
where $\hat{n}_x=\vec{0}$, $\hat{n}_y=\hat{n}_1$, and $\hat{n}_z=\hat{n}_2$. The choice of unit cell and the definition of the Bravais lattice vectors $\hat{n}_{1(2)}$ are shown in Supplementary Fig.~\ref{fig:app_lattice}.

\begin{table}
	\footnotesize
	%\scriptsize
	\renewcommand\arraystretch{1.3}
	\setlength\tabcolsep{0.15cm}
	\begin{tabular}{c|c|c|c|c|c|c|c|c|c}
		\hline
		\hline
		Phase & $(J_z,D_z)$ & $Q^{xy}$ & $\Delta^{zz}_z$ & $\Delta^{xx}_z$ & $\Delta^{yy}_z$ & $\Delta^{zz}_x$& $\Delta^{xx}_x$ & $\Delta^{yy}_x$ & $\Delta^{zx}_x$ \\
		\hline
		Isotropic &$(1,0)$     & $0$         & $0.760$ & $-0.093$ & $0.525$ &  $0.120$ & $0.547$ & $0.525$  & $-0.369$ \\
		$A_0$ &$(0.6,0)$ & $-0.633$ & $0.150$ & $-0.0016$ & $0.057$ &  $0.540$ & $0.381$ & $-0.458$  & $-0.390$ \\
		$A_0$ &$(0.2,0)$ & $-0.644$ & $0.060$ & $-0.0005$ & $0.025$ &  $0.550$ & $0.373$ & $-0.455$  & $-0.390$ \\
		$B$ & $(1.2,0)$ & $ 0.891$ & $0.9997$ & $-4\times{}10^{-5}$ & $5\times{}10^{-5}$ &  $0.005$ & $0.316$ & $-0.316$  & $-0.011$ \\
		$B$ & $(1.6,0)$ & $0.914$ & $0.9999$ & $-1\times{}10^{-5}$ & $2\times{}10^{-5}$ &  $0.003$ & $0.283$ & $-0.283$  & $-0.007$ \\	
		$A_0$ & $(1,1)$ & $-0.860$ & $0.1885$ & $-0.0014$ & $0.0349$ &  $0.5691$ & $0.2146$ & $-0.2627$  & $-0.2886$ \\	
		$A_z$ & $(8,4)$ & $-0.949$ & $0.8365$ & $-8\times{}10^{-4}$ & $1.5\times{}10^{-3}$ &  $0.2377$ & $0.1395$ & $-0.1404$  & $-0.1714$ \\	
		$B$ & $(4,1)$ & $0.9663$ & $1-4\times{}10^{-6}$ & $-3\times{}10^{-7}$ & $3\times{}10^{-7}$ &  $0.00048$ & $0.1813$ & $-0.1813$  & $-0.0014$ \\			
		$A_0$ &$(1,\infty)$ & $-1$ & $0.1807$ & $0$ & $0$ &  $0.627$ & $0$ & $0$  & $0$ \\	
		$A_0$ & $(4,\infty)$ & $-1$ & $0.5248$ & $0$ & $0$ &  $0.5248$ & $0$ & $0$  & $0$ \\	
		$A_z$ & $(8,\infty)$ & $-1$ & $0.9104$ & $0$ & $0$ &  $0.2155$ & $0$ & $0$  & $0$ \\	
		\hline
		\hline
	\end{tabular}
	\caption{The self-consistent solutions for some typical values of $J_z$ and $D_z$. Here $J_x=J_y=1$ and a zero-flux configuration of $\{u\}=1$ is chosen. Notice that  the self-consistent solution for effective isotropic S=1/2 Kitaev spin liquid at $(J_z=4,D_z\rightarrow{}\infty)$ is equivalent to that given in Ref.~\cite{You2012} up to a factor of $-2$, where the minus sign is caused by the antiferromagnetic couplings $J_a>0$ and the factor of $2$ is caused by the normalization condition of Majorana fermions used here, {\em e.g.}, $(\theta^a_i)^2=(\eta^a_i)^2=1$.}\label{tab:app_solutions}
\end{table}

\begin{figure}[ht]
	\centering
	\includegraphics[width=0.9\linewidth]{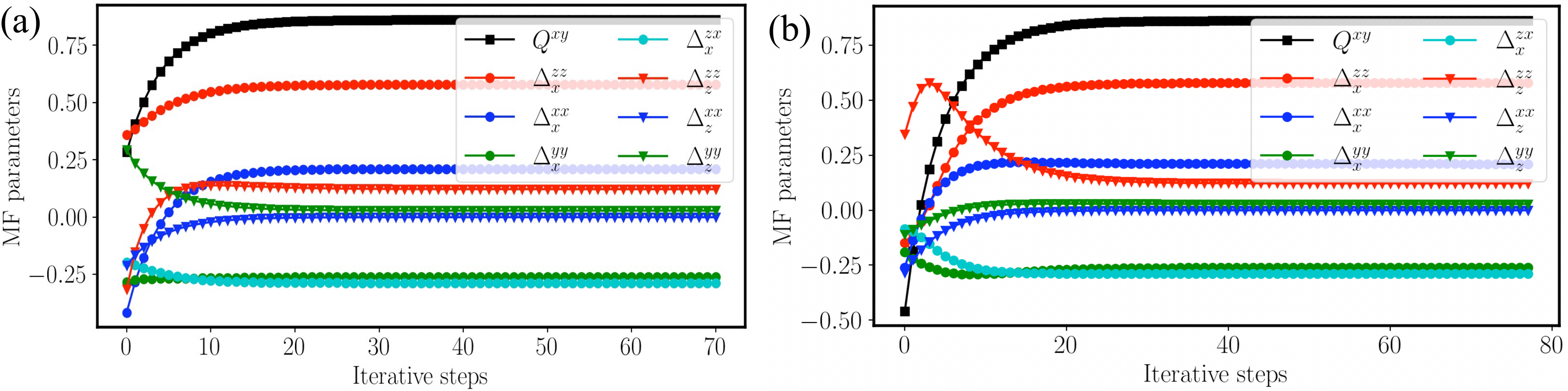}
	\caption{The mean-field parameters as a function of iterative steps, where $J_x=J_y=1$, $J_z=0.6$ and $D_z=1$. Different randomly generated parameters are selected to initialize the iterations in (a) and (b).\label{fig:mfparameters}}
\end{figure}

An iterative scheme has been employed to find the self-consistent solutions to the mean-field Hamiltonian introduced in the main text, where different randomly-generated parameters are selected to initialize the iterations. If multiple inequivalent solutions occurred, we will select the solution with the lowest MF ground-state energy.
The self-consistent solutions for some typical values of $J_z$ and $D_z$ are shown in Supplementary Table~\ref{tab:app_solutions}.
We also show the results of the mean-field parameter as a function of iterative steps in our iterative self-consistent calculations in Supplementary Figs.~\ref{fig:mfparameters}(a) and (b), where two different randomly generated parameters converge to the same solution.

\subsection*{Supplementary Note 2: Spin quadrupolar parameter}

The spin quadrupolar parameter, which is distinguished from magnetic order, is time-reversal invariant. This order usually does not exist in the $S=1/2$ systems because a product of arbitrary two spin-1/2 operators is still a spin-1/2 operator or a trivial identity matrix. While for higher spin systems, the product of two spin operators gives rise to nontrivial spin quadrupolar operators which generally can support the spin quadrupolar parameters. 
A simple example of spin quadrupolar parameter for $S=1$ systems can be found in Ref.~\cite{Lauchli2006}. %

\begin{figure}[ht]
	\centering
	\includegraphics[width=0.8\linewidth]{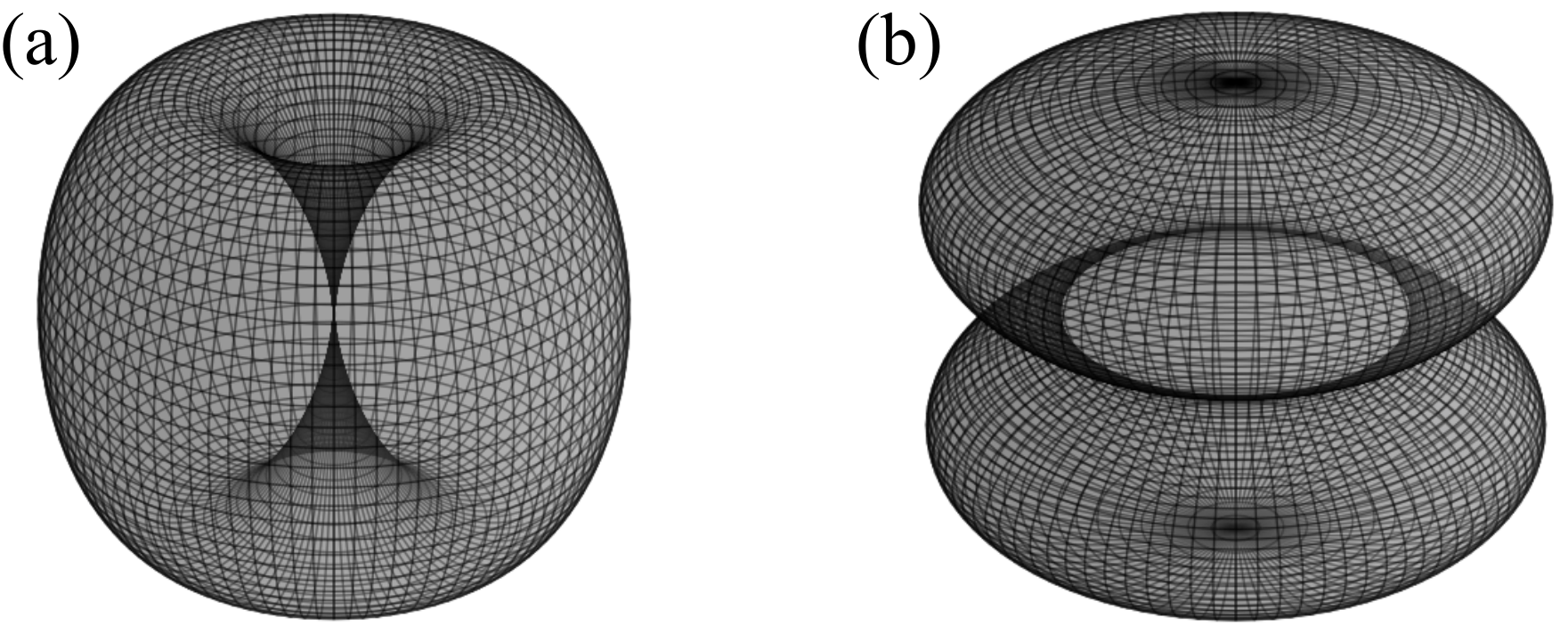}
	\caption{Probabilities of spin fluctuations $P(Q^{xy}, \hat{n})$ [see definition in Eq.~\eqref{eq:spinfluct}] for (a) $Q^{xy}=-1$  and (b) $Q^{xy}=1$.\label{fig:SQO}}
\end{figure}

The Z$_2$ quantum spin liquid state $|\Psi\rangle$ shown in the phase diagram in the main text generally coexists with a spin quadrupolar parameter, {\em e.g.},
\begin{align}
	\langle\Psi| S_j^a|\Psi\rangle&=0,
	\nonumber \\
	\langle\Psi|(S_j^x)^2-(S_j^y)^2|\Psi\rangle&=0,
	\label{eq:SQ}\\
	\langle\Psi|(S_j^z)^2|\Psi\rangle-5/4&=Q^{xy},
	\nonumber
\end{align}
where the last line in Eq.~\eqref{eq:SQ} is the basic definition of the spin quadrupolar parameter in the main text.
Note that these properties in Eq.~\eqref{eq:SQ} cannot be resolved by a single product state. %Since 

The spin quadrupolar parameter usually is illustrated by the probabilities of spin fluctuations~\cite{Lauchli2006}. %$|\langle{S_j(\hat{n})}|Q\rangle|^2$, where $|Q\rangle$ is a local state exhibiting a spin quadrupolar parameter and $|S_j(\hat{n})\rangle$ is an $S=1$ spin coherent state pointing in direction $\hat{n}$ with $$
Since $|\Psi\rangle$ is a many-body state, for illustration purposes we utilize the probabilities of spin fluctuations between S=3/2 spin coherent states $|S(\hat{n})_j\rangle$ and the reduced density matrix $\rho_j(Q^{xy})$ for site $j$:
\begin{equation}
	P\left(Q^{xy},\hat{n}\right)=\mbox{Tr}\left[|{S_j(\hat{n})}\rangle\langle{S_j(\hat{n})}|~\rho_j(Q^{xy})\right],\label{eq:spinfluct}
\end{equation} 
where $|S_j(\hat{n})\rangle$ with $\left(\hat{n}\cdot{\vec{S}}_j\right)|{S_j(\hat{n})}\rangle=S|{S_j(\hat{n})}\rangle$ $\hat{n}$ is a spin coherent state pointing to the direction of $\hat{n}$. 
The probabilities of spin fluctuations for $Q^{xy}=-1$ and $Q^{xy}=+1$ are shown in Supplementary Figs.~\ref{fig:SQO}(a) and (b), respectively. 
Here the reduced density matrix $\rho_j(Q^{xy})$, which can resolve Eq.~\eqref{eq:SQ}, is obtained by exact diagonalization on a $2\times2$ torus.

%\bibliography{S3_2KitaevSO6MF.bib}

\end{widetext}

\end{document}